\documentclass[ApJ, twocolumn]{aastex631}

\usepackage{graphicx}
\usepackage{amsmath}
\usepackage{color}
\usepackage[normalem]{ulem}

\newcommand{\chieff}{\chi_\mathrm{eff}}
\newcommand{\chidif}{\chi_\mathrm{diff}}

\newcommand{\SPA}{School of Physics and Astronomy, Monash University, Clayton VIC 3800, Australia}
\newcommand{\OzGravMonash}{OzGrav: The ARC Centre of Excellence for Gravitational Wave Discovery, Clayton VIC 3800, Australia}

\shorttitle{Evidence for a correlation in $(q,\chieff)$}
\shortauthors{Adamcewicz et al.}

\begin{document}

\title{
Evidence for a correlation between binary black hole mass ratio and black-hole spins}

\author{Christian Adamcewicz}
\email{christian.adamcewicz@monash.edu}
\affiliation{\SPA}
\affiliation{\OzGravMonash}

\author{Paul D. Lasky}
\affiliation{\SPA}
\affiliation{\OzGravMonash}

\author{Eric Thrane}
\affiliation{\SPA}
\affiliation{\OzGravMonash}

\begin{abstract}
The astrophysical origins of the binary black hole systems seen with gravitational waves are still not well understood.
However, features in the distribution of black-hole masses, spins, redshifts, and eccentricities provide clues into how these systems form.
Much has been learned by investigating these distributions one parameter at a time.
However, we can extract additional information by studying the covariance between pairs of parameters.
Previous work has shown preliminary support for an anti-correlation between mass ratio $q \equiv m_2/m_1$ and effective inspiral spin $\chieff$ in the binary black hole population.
In this study, we test for the existence of this anti-correlation using updated data from the third gravitational wave transient catalogue (GWTC-3) and improve our copula-based framework to employ a more robust model for black-hole spins.
We find evidence for an anti-correlation in $(q, \chieff)$ with 99.7\% credibility.
This may imply high common-envelope efficiencies, stages of super-Eddington accretion, or a tendency for binary black hole systems to undergo mass-ratio reversal during isolated evolution.
Covariance in $(q,\chieff)$ may also be used to investigate the physics of tidal spin-up as well as the properties of binary-black-hole-forming active galactic nuclei.
\end{abstract}


\section{Introduction} \label{sec:intro}
There are a number of theorised formation mechanisms for merging binary black hole (BBH) systems \citep[see reviews by][]{Mapelli_2018, Mandel_2022, Spera_2022}.
These scenarios typically fall into one of two broad categories.
The first is dynamical formation, in which black holes are brought together through interactions with other black holes to form merging binaries in dense stellar environments.
The second is isolated (or ``field'') formation, in which stellar black-hole progenitors already in a binary evolve into black holes together and precede to merge afterwards.
However, it remains unclear the relative proportions these channels contribute to the BBH population observed in gravitational waves \citep[see][]{GWTC3} by the LIGO-Virgo-KAGRA collaboration \citep[LVK;][]{LIGO, Virgo, KAGRA}.

This problem of formation is investigated through population studies, which probe the distributions of BBH masses, spins, redshift, and eccentricity for features that may point to a given formation channel \cite[see, for example][and the references therein]{GWTC1_analysis, GWTC2_analysis, GWTC3_analysis, Farr_2018, Baibhav_2020, Wong_2021, Zevin_2021, Bouffanais_2021, cheng_2023}.
Most of these works have focused on one parameter at a time.
While such single-parameter studies are useful, one may gain additional insights by considering the interplay between parameters.
For example, certain isolated formation channels imply metallicity-driven correlations between redshift and BBH mass \citep{Neijssel_2019, Bavera_2020, van_Son_2022}, as well as redshift and spins \citep{Neijssel_2019, Bavera_2022, van_Son_2022, Mapelli_2022}.
Meanwhile, covariance between BBH mass ratio and spins can arise due to various phenomena including common-envelope physics \citep{Bavera_2021, Zevin_2022}, super-Eddington accretion during binary evolution \citep{Bavera_2021, Zevin_2022}, mass-ratio reversal \citep{Broekgaarden_2022}, tidal spin-up of BBH progenitors \citep{Qin_2018, Ma_2023}, hierarchical mergers \citep{Gerosa_2017, Fishbach_2017, Rodriguez_2019, Doctor_2020}, or formation in active galactic nuclei more generally \citep{Mckernan_2022, agn_avi}.

A number of previous studies have looked for covariance between various parameters describing the BBH population \citep{Safarzadeh_2020, Callister_2021, Fishbach_2021, Hoy_2022, Biscoveanu_2022, Franciolini_2022, Tiwari_2022, Adamcewicz_2022, Belczynski_2022, Vitale_2022, Wang_2022, Baibhav_2023, Ray_2023}.
In particular, \cite{Callister_2021} argued that the data support an anti-correlation between mass ratio
\begin{equation}
    q \equiv \frac{m_2}{m_1},
\end{equation}
and effective inspiral spin \citep{Damour_2001}
\begin{equation} \label{eq:chieff}
    \chieff \equiv \frac{\chi_1 \cos t_1 + q \chi_2 \cos t_2}{1 + q}.
\end{equation}
Here, $m_1$ and $m_2$ are the heavier and lighter component masses respectively, $\chi_{1,2}$ are their corresponding dimensionless spin magnitudes, and $\cos t_{1,2}$ are (the cosines of) the tilts of each spin vector measured about the orbital angular momentum.
Using BBH events from the second gravitational wave transient catalogue \citep[GWTC-2;][]{GWTC2}, \cite{Callister_2021} measured such an anti-correlation with 98.7\% credibility.\footnote{
That is, the highest posterior density interval (HDPI) credible interval excludes the no-correlation hypothesis with 98.7\% credibility.
}
Using the same methodology, \cite{GWTC3_analysis} updated this measurement using additional events from the third gravitational wave transient catalogue \citep[GWTC-3;][]{GWTC3}, bringing the credibility for anti-correlation in $(q,\chieff)$ to 97.5\%.

In our previous study \citep{Adamcewicz_2022}, we suggested that investigations like the one carried out by \cite{Callister_2021} \citep[and consequently][]{GWTC3_analysis} are in-principle susceptible to a subtle statistical effect: if the marginal distribution of one parameter implicitly varies with the correlation parameter, then it is difficult to know if the apparent correlation is genuinely present in the data, or if it only seems so because of an improved fit of the marginal distribution.
In other words, one may ask: does LVK data support an anti-correlation between $(q, \chieff)$, or does the data merely prefer the marginal distribution for $\chieff$ implied by anti-correlation?

\cite{Adamcewicz_2022} addresses this concern using a copula density function \citep{Sklar_1996}: a statistical tool that imposes a variable correlation $\kappa$ between two parameters but keeps their corresponding marginal distributions fixed.
\cite{Adamcewicz_2022} corroborates the results of \cite{Callister_2021}, inferring an anti-correlation in $(q,\chieff)$ with 98.7\% credibility with GWTC-2 events and demonstrating that the evidence for anti-correlation is not due merely to a better model for $\chieff$.\footnote{
By coincidence, the analyses in \cite{Callister_2021} and \cite{Adamcewicz_2022} result in the same 98.7\% credibility for anti-correlation, despite utilising different statistical frameworks with different assumptions.
}\footnote{\cite{Adamcewicz_2022} also shows that the anti-correlation between $(q, \chieff)$ is not an inadvertent consequence of implicit assumptions about the distribution of physical spin parameters $\{\chi_{1,2}, \cos t_{1,2}\}$ \citep[see, for example][]{GWTC3_analysis}.
}

However, \cite{Adamcewicz_2022} does not provide an extensive investigation of model misspecification to determine if the anti-correlation can be explained away as arising due to some assumption in our models.\footnote{For an overview of misspecification in gravitational-wave astronomy, see \cite{wmf}.}
\cite{Callister_2021} and \cite{Adamcewicz_2022} mention the potential for the inferred anti-correlation to be a result of two separate, independent sub-populations, neither of which exhibit the observed anti-correlation, but which produce a global anti-correlation via the amalgamation paradox \citep[see][]{Simpson, Pearson_1899, Yule_1903}.

Perhaps a more pressing concern, however, is that the measurement could be affected by an unmodeled sub-population of non-spinning $\chieff=0$ systems.
Given the inconclusive evidence for a non-spinning sub-population of BBH systems found in the literature \citep{Tong_2022, Callister_2022, Mould_2022, Galaudage_2021, Roulet_2021}, it is important to consider how such a feature may influence the inferred shape of the distribution of mass ratio and $\chieff$.

In this work, we aim to more thoroughly investigate the purported $(q, \chieff)$ anti-correlation from \cite{Callister_2021}, \cite{Adamcewicz_2022}, and \cite{GWTC3_analysis}.
We begin in Section~\ref{sec:model} by constructing a more comprehensive population model than the one used in \cite{Adamcewicz_2022}, which allows for a non-spinning sub-population.
We then fit this model to BBH data from the third gravitational wave transient catalogue \citep[GWTC-3;][]{GWTC3} and present the inferred $(q, \chieff)$ correlation, as well as other population properties, in Section~\ref{sec:results}.
We conclude in Section~\ref{sec:discussion}, exploring the implications of these results.

\section{Model} \label{sec:model}

\subsection{Effective coordinates versus physical coordinates}
Before we define our model for binary black hole spin, it is useful to distinguish between two different coordinate systems.
Binary black hole spins are defined by six parameters, three for each black hole: the spin magnitudes of each black hole $\chi_{1,2}$, (the cosines of) the spin tilts of each black hole $\cos t_{1,2}$, and the the azimuthal angles of each component $\phi_{1,2}$.
While the distribution of spin magnitudes and tilts is believed to vary depending on the formation channel, the azimuthal angles are very likely to be uniformly distributed no matter the formation scenario.\footnote{
Although, it is unclear how spin-orbital resonances \citep[see][]{Gerosa_2014} may affect this hypothesis.
}
Hence, models for BBH spin are fully characterised by four dimensions: $\{\chi_{1,2}, \cos t_{1,2}\}$.

Both \cite{Callister_2021} and \cite{Adamcewicz_2022} model black hole spin using only the effective inspiral spin $\chieff$.
The advantage of using this parameter is that $\chieff$ is an approximate constant of motion \citep{Damour_2001} and is relatively well measured from gravitational-wave signals.
In contrast, the spin tilts $\cos t_{1,2}$ evolve as the binary precesses, and are relatively poorly measured.
However, modelling only the distribution of $\chieff$ does not fully determine the distribution of binary black hole spins.
The distributions of the remaining three spin dimensions are defined \textit{implicitly} based on the priors used for initial parameter estimation \citep[see][]{Thrane_2019}, which may not be physically well-motivated.

To address this, we explicitly model $\chieff$ as well as three additional spin parameters so that the distribution of black-hole spins is fully specified.
Following \cite{Roulet_2021}, we define
\begin{equation}
    \chidif \equiv \frac{q \chi_1 \cos t_1 - \chi_2 \cos t_2}{1 + q},
\end{equation}
as well as the in-plane spin magnitudes of each black hole
\begin{equation}
    \rho_i \equiv \chi_i \sqrt{1 - \cos t_i^2},
\end{equation}
where $i=\{1,2\}$.
We refer to the set of $\{\chieff, \chidif, \rho_{1,2}\}$ as the \textit{effective coordinate system}, which can be contrasted with the \textit{physical coordinates} $\{\chi_{1,2}, \cos t_{1,2}\}$.
Either set can be used to fully specify the distribution of BBH spin.

\subsection{Effective spin coordinate model}\label{sec:spin_models}
\begin{figure*}
    \centering
    \includegraphics[width=\textwidth]{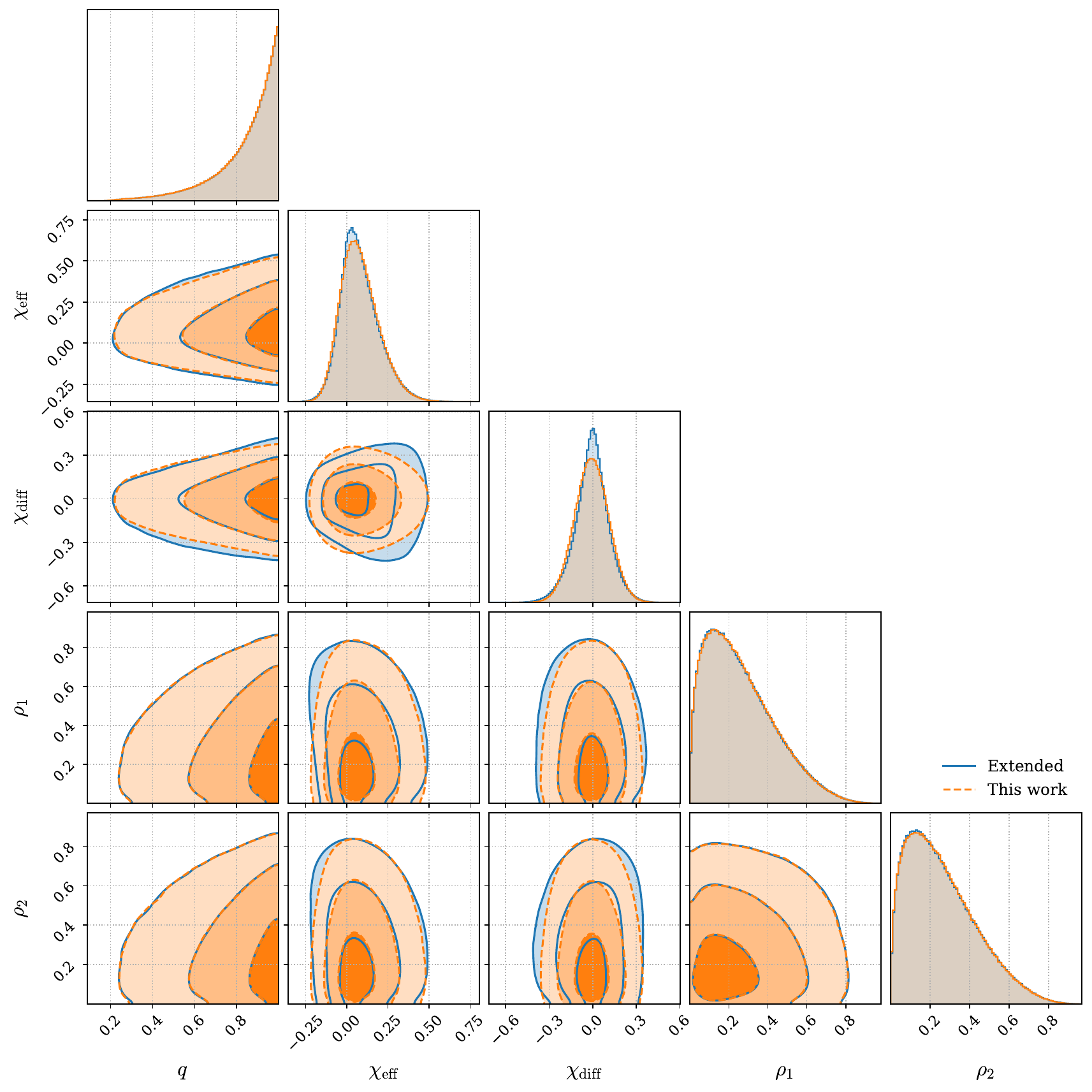}
    \caption{Comparison between the \textsc{Extended} population model from \cite{Tong_2022} (blue), and this work (orange). In this demonstration, the hyper-parameters for our model are chosen to resemble the \textsc{Extended} model distributions in blue. The correlation parameter $\kappa$ is set to zero in this example. Varying $\kappa$ changes the degree of correlation in the $(q,\chieff)$ plane, but leaves every other panel unchanged. The non-spinning sub-population is also excluded from this demonstration.}
    \label{fig:extended_implications}
\end{figure*}

In this subsection we describe a phenomenological model for the marginal distributions of the four effective spin coordinates.
Our population model is inspired by the so-called \textsc{Extended} model recently used in \cite{Galaudage_2021} and \cite{Tong_2022}, which has the flexibility for two sub-populations: one with spinning black holes, and one with non-spinning black holes.
For the moment, we set aside the (trivially defined) non-spinning sub-population (which we revisit in Section~\ref{sec:full_model}), and describe our model for spinning black holes.
Since the \textsc{Extended} model employs physical coordinates, our first step is to define an effective-coordinate model that is phenomenologically similar.
The distributions of effective spin parameters implied by the \textsc{Extended} model are shown in Fig.~\ref{fig:extended_implications} (blue).
We parameterise the distributions of $\chieff$, $\chidif$ and $\rho_{1,2}$, with the intent of capturing the salient features from the \textsc{Extended} model.
Our chosen effective spin models are over-plotted in Fig.~\ref{fig:extended_implications} (orange).
A summary of the models is provided in Table~\ref{tab:priors}.

We model $\chieff$ and $\chidif$ using a skewed normal distribution truncated on the interval $[-1,1]$:
\begin{align}
    \pi(\chieff|\Lambda) = & \ \mathcal{N}_s(\chieff|\xi_\mathrm{eff},\omega_\mathrm{eff},a_\mathrm{eff}), \\
    \pi(\chidif|\Lambda) = & \ \mathcal{N}_s(\chidif|\xi_\mathrm{diff},\omega_\mathrm{diff},a_\mathrm{diff}) ,
\end{align}
where $\Lambda$ denotes the set of all hyper-parameters controlling the shape of the population distributions.
The (truncated) skewed normal distribution is
\begin{align}\label{eq:skew_norm}
    \mathcal{N}_s(x|\xi,\omega,a) \propto & \ 
    \mathcal{N}(x|\xi, \omega) 
    \left[1 + \mathrm{erf}\left(a \frac{x - \xi}{\omega\sqrt{2}}\right)\right] \times \nonumber\\
    & \Theta(x+1) \Theta(1-x) .
\end{align}
Here, $\mathcal{N}(x|\xi, \omega)$ is a normal distribution with mean $\xi$ and width $\omega$.
Meanwhile, $a$ controls the skew, and $\Theta$ is a Heaviside step function.
The mean of ${\cal N}$ is not the mean of ${\cal N}_s$ and so one should think of $\xi$, $\omega$, and $a$ as shape parameters.
In Appendix~\ref{sec:conversions}, we derive the mean $\mu$, variance $\sigma^2$, and skewness $\gamma$ of the ${\cal N}_s$ distribution in terms of these three shape parameters.
This allows us to rewrite our population models for $\chieff$ and $\chidif$ in their final form:
\begin{align}
    \pi(\chieff | & \mu_\mathrm{eff}, \sigma_\mathrm{eff}^2, \gamma_\mathrm{eff}), \\
    \pi(\chidif | & \mu_\mathrm{diff}, \sigma_\mathrm{diff}^2, \gamma_\mathrm{diff}) .
\end{align}

We model $\rho_{1,2}$ using beta distributions:
\begin{equation}\label{eq:rho_dist}
    \pi(\rho_1, \rho_2|\Lambda) = \mathrm{Beta}(\rho_1|\mu_\rho, \sigma_\rho^2) \mathrm{Beta}(\rho_2|\mu_\rho, \sigma_\rho^2),
\end{equation}
where $\mu_\rho$ is the Beta distribution mean and $\sigma_\rho^2$ is the variance.
Comparing the blue and orange intervals in Fig.~\ref{fig:extended_implications}, we observe that our new effective-coordinate parameterisation produces similar distributions to the ones implied by the \textsc{Extended} model.

\subsection{Correlating $q$ and $\chieff$}
The next step is to combine our effective spin coordinate model with a model for black-hole masses.
We take primary mass $m_1$ and mass ratio $q$ to be distributed according to the \textsc{Power-Law + Peak} mass model from \cite{Talbot_2018}.
For completeness, we define this model in Appendix~\ref{sec:mass_model}.

With all of the marginal distributions defined, we employ a Frank copula density function \citep{frank} in order to introduce a correlation between $q$ and $\chieff$:
\begin{equation}\label{eq:frank_copula}
    \pi_c(u,v|\kappa) = \frac{\kappa e^{\kappa(u+v)}(e^\kappa - 1)}{\left(e^\kappa - e^{\kappa u} - e^{\kappa v} + e^{\kappa(u + v)}\right)^2}.
\end{equation}
Here, $u$ and $v$ are copula coordinates determined by the marginal distributions of $q$ and $\chieff$ respectively:
\begin{align}
    \label{eq:u}
    u(q|\Lambda) &= \int_{m_{\min}/m_1}^q dq' \pi(q'|\Lambda), \\
    \label{eq:v}
    v(\chieff|\Lambda) &= \int_{-1}^{\chieff} d\chieff' \pi(\chieff'|\Lambda).
\end{align}
For a Frank copula, $\kappa < 0$ implies correlation, $\kappa > 0$ implies anti-correlation, and $\kappa=0$ implies no correlation.
For a more detailed introduction to copulas in population studies, see \cite{Adamcewicz_2022}.

\subsection{Final touches} \label{sec:full_model}
For spinning black holes, the joint distribution of mass and spin is
\begin{align} \label{eq:spin_model}
    \pi_s(\theta | \Lambda, \kappa) = & \ \pi(m_1, q | \Lambda)
    \pi(\chieff, \chidif, \rho_{1,2} | \Lambda) \times \nonumber\\
    & \ \pi_c \Big(u(q|\Lambda), v(\chieff|\Lambda) | \kappa \Big),
\end{align}
where we denote the full set of parameters $\theta = \{m_1, q, \chieff, \chidif, \rho_{1,2}\}$.
Here, $\pi(m_1,q|\Lambda)$ is the \textsc{Power-Law + Peak} mass model described in Appendix~\ref{sec:mass_model}, and $\pi(\chieff,\chidif,\rho_{1,2}|\Lambda)$ is the product of the one-dimensional spin distributions defined in Section~\ref{sec:spin_models}.
Note that while the distribution of effective spin coordinates factorizes, the distribution of physical spin coordinates does not.

Our model includes a second sub-population of non-spinning black holes, described as
\begin{equation}\label{eq:no_spin_model}
    \pi_0(\theta|\Lambda) =
    \pi(m_1,q|\Lambda) \delta(\chieff) \delta(\chidif) \delta(\rho_1) \delta(\rho_2),
\end{equation}
where $\delta$ is a Dirac delta function, and $\pi(m_1,q|\Lambda)$ is the same \textsc{Power-Law + Peak} mass model used for the spinning sub-population.
Putting everything together, the full model is
\begin{equation}
    \pi(\theta|\Lambda,\kappa,\lambda_0) = \lambda_0 \pi_0(\theta|\Lambda) + (1-\lambda_0)\pi_s(\theta|\Lambda,\kappa).
\end{equation}
Here, $\lambda_0$ is a mixing fraction corresponding to the fraction of binaries with non-spinning black holes.
The priors for each hyper-parameter in $\Lambda$, along with $\kappa$ and $\lambda_0$, are listed in Table~\ref{tab:priors}.

\section{Method}

\begin{table*}
    \centering
    \begin{tabular}{|c l l|}
        \hline
        \multicolumn{3}{|c|}{$m_1$ -- \textsc{Power-Law + Peak}}\\
        \hline
        $\alpha$ & $\mathcal{U}(-3, 6)$ & 
        Spectral-index of power-law component\\
        \hline
        $\mu_m$ & $\mathcal{U}(20\text{M}_\odot, 50\text{M}_\odot)$ & 
        Mean of Gaussian component\\
        \hline
        $\sigma_m$ & $\mathcal{U}(1\text{M}_\odot, 10\text{M}_\odot)$ & 
        Width of Gaussian component\\
        \hline
        $\lambda_m$ & $\mathcal{U}(0,1)$ & 
        Fraction of masses in Gaussian component\\
        \hline
        $m_{\min}$ & $\mathcal{U}(2\text{M}_\odot, 10\text{M}_\odot)$ & 
        Minimum allowed mass\\
        \hline
        $m_{\max}$ & $\mathcal{U}(60\text{M}_\odot, 100\text{M}_\odot)$ & 
        Maximum allowed mass\\
        \hline
        $\delta_m$ & $\mathcal{U}(0, 10\text{M}_\odot)$ & 
        Smoothing length for minimum mass\\
        \hline
        \hline
        \multicolumn{3}{|c|}{$q$ -- \textsc{Power-Law}}\\
        \hline
        $\beta_q$ & $\mathcal{U}(-3, 6)$ & 
        Spectral-index of power-law\\
        \hline
        \hline
        \multicolumn{3}{|c|}{$\chieff$ -- \textsc{Skewed Gaussian}}\\
        \hline
        $\mu_\mathrm{eff}$ & $\mathcal{U}(-1, 1)$ & 
        Mean of distribution\\
        \hline
        $\sigma_\mathrm{eff}^2$ & $\mathcal{U}(10^{-3},0.25)$ & 
        Variance of distribution\\
        \hline
        $\gamma_\mathrm{eff}$ & $\mathcal{U}(-0.99,0.99)$ & 
        Skewness of distribution\\
        \hline
        \hline
        \multicolumn{3}{|c|}{$\chidif$ -- \textsc{Skewed Gaussian}}\\
        \hline
        $\mu_\mathrm{diff}$ & $\mathcal{U}(-1, 1)$ & 
        Mean of distribution\\
        \hline
        $\sigma_\mathrm{diff}^2$ & $\mathcal{U}(10^{-3},0.25)$ & 
        Variance of distribution\\
        \hline
        $\gamma_\mathrm{diff}$ & $\mathcal{U}(-0.99,0.99)$ & 
        Skewness of distribution\\
        \hline
        \hline
        \multicolumn{3}{|c|}{$\rho_{1,2}$ -- \textsc{Beta}}\\
        \hline
        $\mu_{\rho}$ & $\mathcal{U}(0,1)$ & 
        Mean of both distributions\\
        \hline
        $\sigma_{\rho}^2$ & $\mathcal{U}(10^{-3},0.25)$ & 
        Variance of both distributions\\
        \hline
        \hline
        \multicolumn{3}{|c|}{\textsc{Other}}\\
        \hline
        $\kappa$ & $\mathcal{U}(-50,50)$ & 
        Level of correlation between $q$ and $\chieff$\\
        \hline
        $\lambda_0$ & $\mathcal{U}(0,1)$ & 
        Fraction of non-spinning systems\\
        \hline
    \end{tabular}
    \caption{List of models and associated hyper-parameters for each BBH parameter, along with corresponding priors. Here, $\mathcal{U}(a,b)$ indicates a uniform prior on the interval $[a,b]$. We cut the prior on $\gamma_\mathrm{eff(diff)}$ at $\pm 0.99$, because the skewness parameter is only well defined when $a_\mathrm{eff(diff)}(1 + a_\mathrm{eff(diff)}^2)^{-1} \in [-1,1]$. These limits roughly correspond to $\gamma_\mathrm{eff(diff)} \in [-0.99, 0.99]$. Furthermore, we cut the prior on $\kappa$ at $\pm 50$. Beyond these limits, the implied level of correlation plateaus.
    }
    \label{tab:priors}
\end{table*}

We measure the population hyper-parameters using hierarchical Bayesian inference; see Appendix~\ref{sec:inference} and \cite{Thrane_2019}.
We employ the population inference package \texttt{GWPopulation} \citep{Talbot_2019} -- which is built on top of \texttt{Bilby} \citep{bilby,bilby_gwtc1} -- and employ the nested sampler \texttt{DYNSETY} \citep{Speagle_2020}.
We use BBH events from GWTC-3 \citep{GWTC3, zenodo, zenodo_sens} that were considered reliable for inclusion in the LVK O3b population analysis \citep{GWTC3_analysis}.
However, following \cite{Tong_2022}, we exclude the events GW200129\_065458 and GW191109\_010717 (hereby shortened to GW200129 and GW191109 respectively) from our analysis over data-quality concerns.
The former event GW200129 was coincident with a detector glitch.
While parameter estimation on cleaned data shows strong evidence for spin-precession \citep{GWTC3, Hannam_2022}, the results may be sensitive to the glitch subtraction procedure \citep{Payne_2022}, which is still under discussion \citep{Davis_2022, Macas_2023}.
The latter event GW191109 is still under investigation \citep{hui}.
We repeat our analysis both with and without GW191109, bringing the total number of BBH events surveyed to 68 and 67 in each case.
For more details on how parameter estimation is utilised, how our population likelihood is constructed, and how we account for selection biases in our analyses see Appendix~\ref{sec:inference}.

A subtlety arises when building models in effective coordinates; some combinations of $\chieff$, $\chidif$, and $\rho_{1,2}$ can imply unphysical values for spin magnitudes $\chi_{1,2} > 1$.
To correct for this, unphysical samples should be discarded and the population distribution re-normalised accordingly.
We carry out this step in post-processing -- computing a weight for each hyper-posterior sample according to the appropriate normalisation factor.
Before this correction, values of $\chi_{1,2} > 1$ account for $\lesssim 1\%$ of the population for any given draw of $\Lambda$ and $\kappa$ from the hyper-posterior.
As such, the effect of this normalisation is small.

\section{Results} \label{sec:results}

\begin{figure}
    \centering
    \includegraphics[width=\columnwidth]{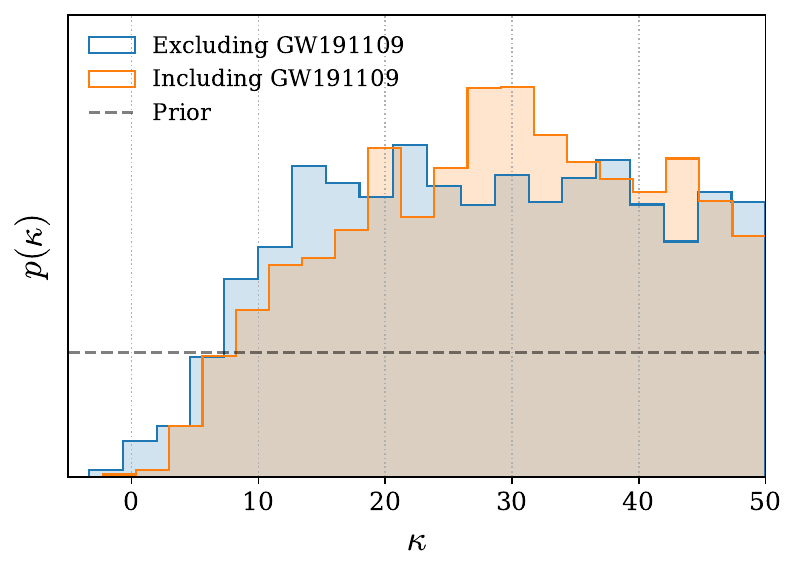}
    \caption{
    Posterior distribution of the $(q,\chieff)$ correlation $\kappa$ for analyses excluding (blue) and including (orange) GW191109. Excluding GW191109, we find 90\% credible intervals of $\kappa = 28^{+20}_{-20}$. When GW191109 is included, we find 90\% credible intervals of $\kappa = 30^{+18}_{-21}$. 
    The prior is indicated with a dashed line, which extends from $\kappa = -50$ to $\kappa = 50$.
}
    \label{fig:kappa_lambda_post}
\end{figure}

In Fig.~\ref{fig:kappa_lambda_post} we plot the posterior distribution for $\kappa$, which measures the degree of correlation between mass ratio and $\chieff$.
We see a strong preference for $\kappa > 0$, corresponding to an anti-correlation in $(q,\chieff)$.
With 90\% credibility, we find $\kappa = 28^{+20}_{-20}$ when GW191109 is excluded, and $\kappa = 30^{+18}_{-21}$ when it is included.
We rule out the uncorrelated hypothesis $(\kappa = 0)$ with 99.7\% credibility.
This number varies only slightly (by $\lesssim 0.1\%$) when we vary the prior bounds on $\kappa$ or include GW191109.

The correlation hypothesis is preferred over the no-correlation hypothesis with a natural log Bayes factor of  $\ln \text{BF} = 2.2$ excluding GW191109 or $\ln \text{BF} = 3.2$ including GW191109.
Moreover, the correlation hypothesis yields a maximum natural log likelihood that is $\Delta \ln \mathcal{L}_{\max} = 6.4$ (9.5) higher than the no-correlation hypothesis including (excluding) GW191109, suggesting that the data are better fit by the correlation model.
The Bayes factor is smaller than the ratio of maximum likelihoods because the correlation hypothesis incurs an Occam’s factor as a result of the additional correlation parameter $\kappa$.
Thus, we report increased evidence for anti-correlation between $(q, \chieff)$ from \cite{Callister_2021} and \cite{Adamcewicz_2022}, both of which analysed only the 44 BBH events from GWTC-2 \citep{GWTC2}, and from \cite{GWTC3_analysis}, which analysed all events from GWTC-3 \citep{GWTC3}.

\begin{figure}
    \centering
    \includegraphics[width=\columnwidth]{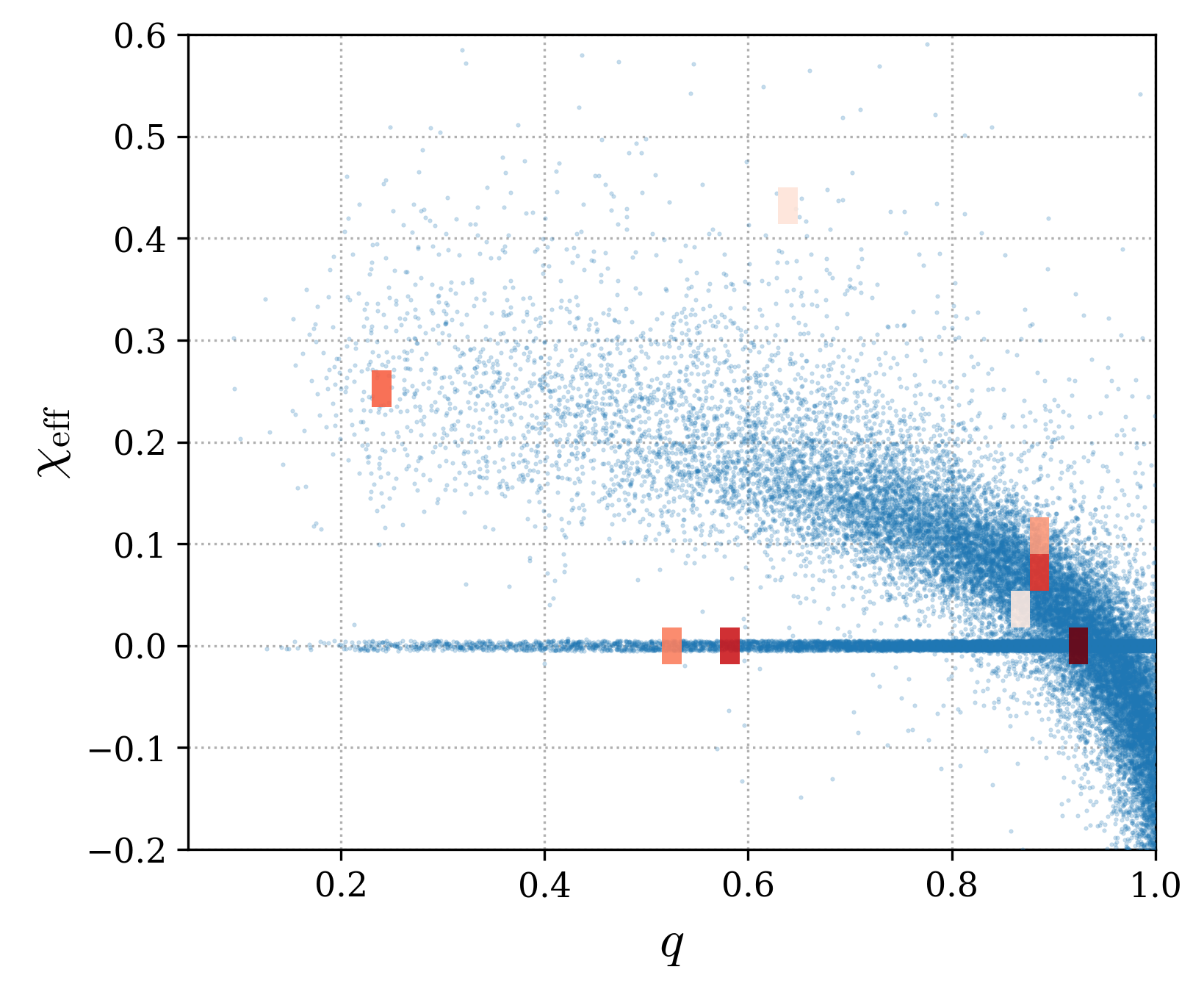}
    \caption{Plot of $3 \times 10^4$ draws from the population predictive distribution for mass ratio and $\chieff$, marginalised over the posterior distributions of the population hyper-parameters and excluding GW191109. In red, is the purely data-driven \sout{$\pi$} from \cite{Payne_2023}, consisting of a set of delta functions that attempt to maximise the population likelihood.
    Darker tones indicate a higher weight, thus higher importance, for the given feature.
    Our model matches qualitative features present in the \sout{$\pi$} result.
    Note that samples drawn from the non-spinning sub-population have been randomly off-set by $\chieff \sim \pm 0.005$ in order to help with the visualisation. 
    }
    \label{fig:q_xeff_pop}
\end{figure}

The anti-correlation in $(q,\chieff)$ is illustrated by the reconstructed population distribution (population predictive distribution), plotted in Fig.~\ref{fig:q_xeff_pop}.
Superimposed over this distribution, is a set of weighted delta functions \sout{$\pi$}, which attempt to maximise the population likelihood, taken from \cite{Payne_2023}.
Essentially, \sout{$\pi$} is a data-driven population model, in which the constituent delta functions latch onto the most important features in the data, indicating where most of the population weight should lie \citep[see][for a more detailed explanation]{Payne_2023}.
Our phenomenological model reproduces the most highly-weighted features.

In Appendix~\ref{sec:copulas}, we explore the impact that changing our choice of copula density function has on our measurement of a correlation $\kappa$.
Regardless of the copula we choose, we find that the anti-correlation in $(q,\chieff)$ is still strongly favoured.

Finally, in Appendix~\ref{sec:further_results}, we share results peripheral to the goal of this work, including posterior distributions on all hyper-parameters $\Lambda$, and full reconstructed population distributions.

\section{Discussion} \label{sec:discussion}
Although we have shown stronger evidence for an anti-correlation in $(q,\chieff)$, a modeler's job is never done: while we have addressed the most obvious sources of model-dependency, additional work should be carried out to investigate other possible sources of misspecification.

The possibility of the observed anti-correlation being a result of the amalgamation paradox (see the discussion in Section~\ref{sec:intro}) remains a point of interest.
\cite{Baibhav_2023}, for instance, show that when modelling field-like and dynamical-like sub-populations in physical spin coordinates, the two sub-populations prefer different distributions of mass ratio, which could drive such an effect.
Although \cite{Baibhav_2023} emphasise that we are unable to distinguish between multiple sub-populations in the marginal distribution of $\chieff$ \citep[a point further exemplified in non-parametric population studies;][]{Callister_2023, Edelman_2023}, it remains possible that more clearly separated sub-populations could be seen in the $q$-$\chieff$ plane.

This point aside, there are multiple astrophysical explanations for the apparent $(q, \chieff)$ anti-correlation that also merit future investigation.
Simulations of mass-ratio reversed field binaries by \cite{Broekgaarden_2022} yield a distribution of $(q,\chieff)$, qualitatively similar to our reconstructed population (Fig.~\ref{fig:q_xeff_pop}).
We speculate that the observed anti-correlation could be an imprint of mass-ratio reversal in field binaries.

On the other hand, \cite{Bavera_2021} and \cite{Zevin_2022} predict a similar anti-correlation in isolated BBH systems that undergo common-envelope evolution with high common-envelope efficiency or stable mass transfer involving super-Eddington accretion.
These scenarios, however, would not be expected to dominate the observed BBH population.

\cite{Ma_2023} predict that tidal spin-up rate \citep[see also][]{Qin_2018, Hu_2022} should scale with $q^2$.
Given that isolated systems have preferentially aligned black hole spins (positive $\chieff$), such a scaling would presumably result in a \textit{positive} correlation in $(q,\chieff)$.
This seems to be in tension with our results.
However, \cite{Ma_2023} also cite a number of caveats that could complicate the relationship between mass ratio and tidal spin-up, which are yet to be explored.

An anti-correlated $(q,\chieff)$ might arise from some dynamical channels.
For example, \cite{Mckernan_2022} show how such a feature may arise due to BBH formation in active galactic nuclei (AGN), provided a number of assumptions about the properties of the AGN in question hold.
Namely, that black holes in the disk of the AGN are both more massive and have orbitally aligned spins, while those outside are lighter and isotropically distributed in spin orientation.
Black holes outside the disk are then assumed to undergo turbulent migration into the AGN disk where they form merging binaries.
Therefore, if it can be verified that this pathway is responsible for the observed anti-correlation in $(q,\chieff)$, it may be possible infer properties of AGN environments using BBH mergers \citep{agn_avi}.

Moving forward, it will be beneficial to construct population models that have the capacity to distinguish between the various scenarios outlined above.
Doing so may provide evidence for, or falsify, the aforementioned hypotheses on the origin of the purported $(q,\chieff)$ anti-correlation -- therefore informing our understanding of BBH formation mechanisms more generally.

\section*{Acknowledgements}
We thank our referee, Maya Fishbach, Charlie Hoy, and Christopher Berry for their helpful comments on this manuscript.
We thank Hui Tong for providing the data for non-spinning BBH systems.
We acknowledge support from the Australian Research Council (ARC) Centre of Excellence CE170100004 and ARC DP230103088.
This material is based upon work supported by NSF's LIGO Laboratory which is a major facility fully funded by the National Science Foundation.
The authors are grateful for computational resources provided by the LIGO Laboratory and supported by National Science Foundation Grants PHY-0757058 and PHY-0823459.

This research has made use of data or software obtained from the Gravitational Wave Open Science Center (gw-openscience.org), a service of LIGO Laboratory, the LIGO Scientific Collaboration, the Virgo Collaboration, and KAGRA. LIGO Laboratory and Advanced LIGO are funded by the United States National Science Foundation (NSF) as well as the Science and Technology Facilities Council (STFC) of the United Kingdom, the Max-Planck-Society (MPS), and the State of Niedersachsen/Germany for support of the construction of Advanced LIGO and construction and operation of the GEO600 detector. Additional support for Advanced LIGO was provided by the Australian Research Council. Virgo is funded, through the European Gravitational Observatory (EGO), by the French Centre National de Recherche Scientifique (CNRS), the Italian Istituto Nazionale di Fisica Nucleare (INFN) and the Dutch Nikhef, with contributions by institutions from Belgium, Germany, Greece, Hungary, Ireland, Japan, Monaco, Poland, Portugal, Spain. The construction and operation of KAGRA are funded by Ministry of Education, Culture, Sports, Science and Technology (MEXT), and Japan Society for the Promotion of Science (JSPS), National Research Foundation (NRF) and Ministry of Science and ICT (MSIT) in Korea, Academia Sinica (AS) and the Ministry of Science and Technology (MoST) in Taiwan.

\appendix

\section{A change of hyper-parameters} \label{sec:conversions}
In equation~\ref{eq:skew_norm}, we define the skew norm distributions in terms of variables $\xi$, $\omega$, and $a$.
However, we work in terms of the means $\mu$, variances $\sigma^2$, and skewnesses $\gamma$, defined by
\begin{align}
    \mu &= \xi + \omega \bar{a} \sqrt{\frac{2}{\pi}}, \\
    \sigma^2 &= \omega^2 \left(1 - \frac{2\bar{a}^2}{\pi}\right), \\
    \gamma &= \frac{(4 - \pi)(\bar{a} \sqrt{2/\pi})^3}{2(1 - 2 \bar{a}^2/\pi)^{3/2}},
\end{align}
where
\begin{equation}
    \bar{a} = \frac{a}{\sqrt{1 + a^2}},
\end{equation}
is defined on the interval $[-1,1]$.

\section{Mass model}\label{sec:mass_model}
We assume primary mass $m_1$ and mass ratio $q$ are distributed according to the \textsc{Power-Law + Peak} mass model from \cite{Talbot_2018}.
Primary mass follows a distribution
\begin{equation}
    \pi(m_1|\Lambda) \propto \bigg(\lambda_m \mathcal{N}(m_1|\mu_m, \sigma_m) + (1 - \lambda_m)\mathcal{P}(m_1|\alpha)\bigg)\Theta(m_{\max}-m_1)S(m_1|m_{\min},\delta_m),
\end{equation}
where $\mathcal{N}(m_1|\mu_m, \sigma_m)$ is a Gaussian distribution with mean $\mu_m$ and width $\sigma_m$, $\lambda_m$ is the fraction of masses in this Gaussian component, and
\begin{align}
    \mathcal{P}(m_1 | \alpha) \propto m_1^\alpha,
\end{align}
is a power-law distribution with spectral-index $\alpha$.
The primary mass $m_1$ is restricted to lie between hyper-parameters $[m_{\min}, m_{\max}]$ using a Heaviside step function $\Theta$ at the upper mass limit and a smoothing function at low mass:
\begin{equation}
    S(m_1|m_{\min},\delta_m) = \left[\exp\left(\frac{\delta_m}{m - m_{\min}} - \frac{\delta_m}{m-m_{\min}-\delta_m}\right) + 1\right]^{-1},
\end{equation}
where $\delta_m$ is the smoothing length.
The mass ratio is then power-law distributed with a dependence on $m_1$:
\begin{equation}
    \pi(q|m_1, \Lambda) \propto \mathcal{P}(q | \beta_q) \Theta(1-q) \Theta(q - m_{\min}/m_1).
\end{equation}

\section{Constructing a likelihood}\label{sec:inference}
Usually, upon constructing a population model, one can infer the population hyper-parameters by inserting it into, then minimising, the likelihood function defined in equation~32 of \cite{Thrane_2019}.
This likelihood function effectively re-weights individual event parameter samples -- which have been obtained using an uninformative fiducial prior -- according to the newly proposed population model.
These re-weighted samples are then combined to form a likelihood.
However, the aforementioned fiducial samples generally do not have sufficient prior volume around $\chi_{1,2}=0$, meaning population-level inferences would be biased away from cases of zero-spin.

To remedy this, we use the methodology from \cite{Galaudage_2021} and \cite{Tong_2022}, in which the likelihood uses two sets of fiducial samples.
One set is produced using the standard uninformative prior on $\chi_{1,2}$, being uniform on $[0,1]$.
The other is produced by setting $\chi_{1,2} = 0$.
We then create a pair of corresponding, event-level likelihoods.
For the non-spinning case, given an event $i$ with associated data $d^i$, we have
\begin{equation}
    \mathcal{L}_0(d^i|\Lambda) = \frac{\mathcal{Z}_0^i}{n} \sum_k^n \frac{\pi_0(\theta_k^i|\Lambda)}{\pi_{\O}(\theta_k^i|\chi_{1,2}=0)},
\end{equation}
where $\mathcal{Z}_0^i$ is the evidence obtained during fiducial sampling for event $i$, assuming $\chi_{1,2}=0$, $n$ is the number of samples re-weighted, $\pi_0(\theta_k^i|\Lambda)$ is the model defined in equation~\ref{eq:no_spin_model} for the non-spinning sub-population, evaluated at a sample $\theta_k^i$, and $\pi_{\O}(\theta_k^i|\chi_{1,2}=0)$ is the fiducial prior used in non-spinning sample generation, evaluated for the same sample.
For the spinning case, we then have the familiar likelihood
\begin{equation}
    \mathcal{L}_s(d^i|\Lambda, \kappa) = \frac{\mathcal{Z}_s^i}{n} \sum_k^n \frac{\pi_s(\theta_k^i|\Lambda,\kappa)}{\pi_{\O}(\theta_k^i)},
\end{equation}
where $\mathcal{Z}_s^i$ is the fiducial evidence for event $i$, assuming a uniform spin prior, $\pi_s(\theta_k^i|\Lambda, \kappa)$ is the spinning sub-population model defined in equation~\ref{eq:spin_model}, evaluated at a sample $\theta_k^i$, and $\pi_{\O}(\theta_k^i)$ is the fiducial prior with uniform spin, evaluated at the same sample.

We can then combine these event-level likelihoods with a mixing fraction $\lambda_0$, and produce a total likelihood by taking the product over all $N$ events
\begin{equation}
    \mathcal{L}(d|\Lambda,\kappa,\lambda_0) = \frac{1}{p_\text{det}(\Lambda)^{N}} \prod_i^N \Big[
    \lambda_0 \mathcal{L}_0(d^i|\Lambda) +
    (1 - \lambda_0) \mathcal{L}_s(d^i|\Lambda, \kappa) 
    \Big].
\end{equation}
Here, the factor of $p_\text{det}(\Lambda)^{-N}$ accounts for mass and redshift-based effects on BBH detection efficiency \citep[see][]{Messenger_2013, Thrane_2019, Mandel_2019, GWTC1_analysis, GWTC2_analysis, GWTC3_analysis}.
We opt to exclude selection effects based on BBH spins as there are a number of complications that arise due to sampling issues about spin magnitudes of $\chi=0$.
These spin-based selection biases are generally thought to be minor \citep{GWTC3_analysis} and should not have a significant impact on our inferences.

\section{Impact of alternative copulas}\label{sec:copulas}
When choosing a copula density function, one must consider the types of correlations that the given function has the flexibility for.
Currently in the literature, there appear to be only three copula density functions that allow for a smooth transition from correlation to anti-correlation by varying $\kappa$.
These are the Frank copula density function \citep{frank}, Gaussian copula density function \citep{gaussian}, and Farlie-Gumbel-Morgenstern (FGM) copula density function \citep{farlie, gumbel, morgenstern}.
Here, we investigate the effect of replacing the Frank copula density function, assumed in Section~\ref{sec:model}, with the two latter copulas.
That is, we replace the function $\pi_c(u,v|\kappa)$ in equation~\ref{eq:spin_model}, and re-run our analyses to get alternative inferences on the correlation $\kappa$.
We do so excluding GW191109 from our dataset.

\begin{figure*}
    \centering
    \includegraphics[width=0.8\textwidth]{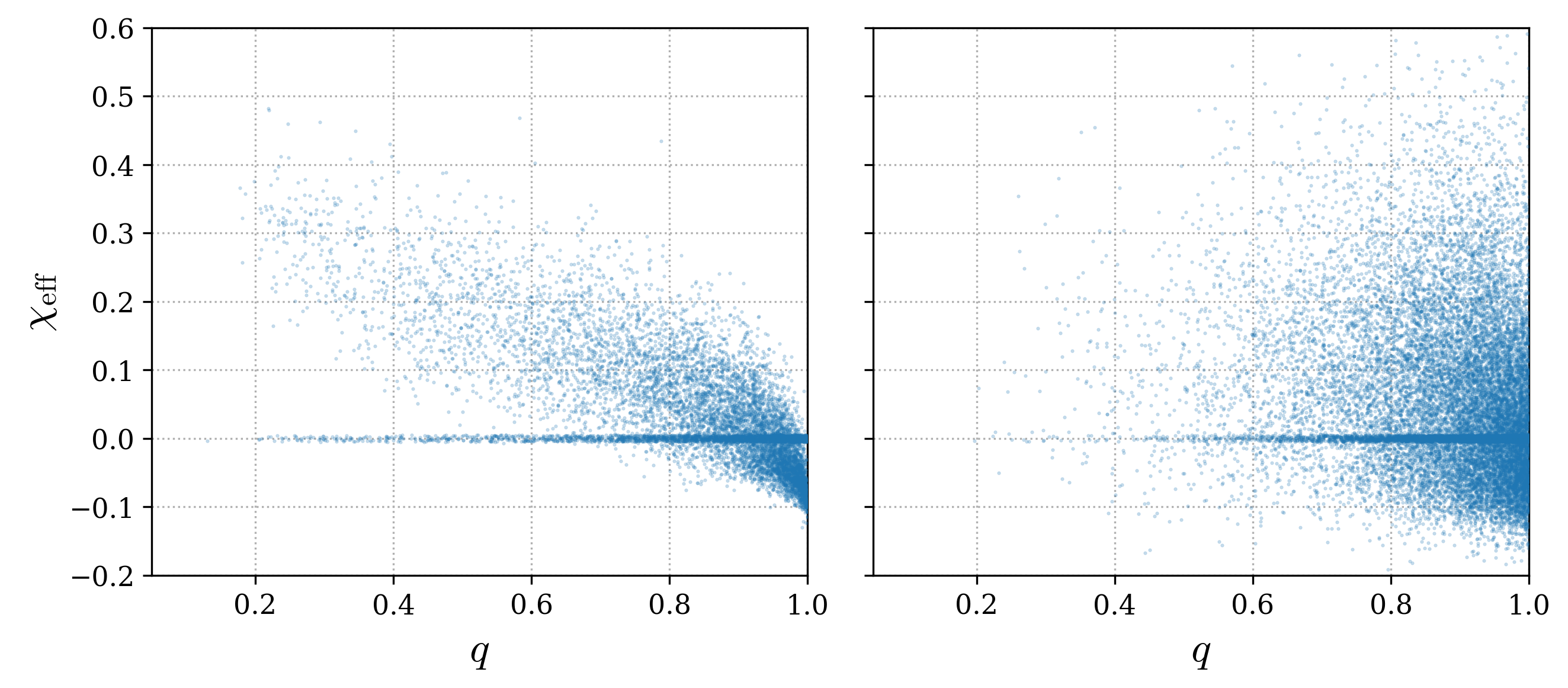}
    \caption{Comparison of $(q,\chieff)$ population distributions for the Gaussian copula model (left) and FGM copula model (right). Plots consist of $2 \times 10^4$ draws from the population predictive distribution for mass ratio and $\chieff$, assuming the maximum posterior probability hyper-parameters. These plots use results from analyses excluding GW191109. Note that samples drawn from the non-spinning sub-population have been randomly off-set by $\chieff \sim \pm 0.005$ in order to help with the visualisation.}
    \label{fig:copula_comp}
\end{figure*}

Firstly, we explore the effects of a Gaussian copula on our inferences.
A Gaussian copula density function can be given as
\begin{equation}
    \pi_c(\bar{u},\bar{v}|\kappa) = 
    \frac{1}{\sqrt{1 - \kappa^2}}
    \exp \left(
    - \frac{\kappa^2(\bar{u}^2 + \bar{v}^2) - 2 \kappa \bar{u} \bar{v}}
    {2 (1 - \kappa^2)}
    \right),
\end{equation}
where
\begin{align}
    \bar{u} &= \sqrt{2} \ \text{erf}^{-1} (2u - 1),\\
    \bar{v} &= \sqrt{2} \ \text{erf}^{-1} (2v - 1),
\end{align}
and $u$ and $v$ are as defined in equations~\ref{eq:u}~and~\ref{eq:v}.
Here, $\kappa$ is restricted to $[-1,1]$, where $\kappa=1$ implies maximum correlation and $\kappa=-1$ implies maximum anti-correlation.\footnote{
Note that this behaviour is opposite to that of the Frank copula density function, in which positive values of $\kappa$ imply anti-correlation.
}
With a Gaussian copula, we find anti-correlation ($\kappa < 0$) with 97.7\% credibility.
Compared to the Frank copula model, the Gaussian copula model is disfavoured by a natural log Bayes factor of only $\ln \text{BF} = 0.1$.
However, the maximum natural log likelihood of the Frank copula model is $\Delta \ln \mathcal{L}_{\max} = 1.2$ higher.
This implies that the Gaussian copula model is favoured over the no-correlation hypothesis by a natural log Bayes factor of $\ln \text{BF} = 2.1$ ($\Delta \ln \mathcal{L}_{\max} = 5.2$).
The $(q,\chieff)$ population distribution for this model is reconstructed using the maximum posterior probability hyper-parameters in the left panel of Fig.~\ref{fig:copula_comp}.
We can see that the Gaussian copula forces the distribution of $\chieff$ to narrow substantially as $q$ approaches one, implying only a small range of $\chieff$ for equal-mass binaries.
This feature may explain the data's slight preference for the Frank copula model.

Next, we experiment with an FGM copula density function, defined as
\begin{equation}
    \pi_c(u,v|\kappa) = 1 + \kappa(1-2u)(1-2v),
\end{equation}
where $\kappa$ is again defined on $[-1,1]$, with $\kappa=1$ and $\kappa=-1$ giving maximum correlation and anti-correlation respectively.
With an FGM copula, we find anti-correlation ($\kappa < 0$) with 79.1\% credibility.
Again comparing to the Frank copula model, the FGM model is disfavoured by a natural log Bayes factor of $\ln \text{BF} = 1.8$ (difference in maximum natural log likelihood of $\Delta \ln \mathcal{L}_{\max} = 2.2$).
Hence, the FGM model is less distinguishable from the uncorrelated model, favoured by a natural log Bayes factor of only $\ln \text{BF} = 0.3$, but with a notable improvement in maximum natural log likelihood $\Delta \ln \mathcal{L}_{\max} = 4.3$.
The $(q,\chieff)$ population distribution for this model is again plotted assuming the maximum posterior probability hyper-parameters in the right panel of Fig.~\ref{fig:copula_comp}.
We can see that correlations manifest themselves more subtly using an FGM copula -- making it unsurprising that this model is less distinguishable from an uncorrelated model.

\section{Full posterior and population distributions}\label{sec:further_results}
Here, we share results that are not relevant to the key question of this study (are mass ratio and $\chieff$ correlated?), but may still be of interest for some readers.

\begin{figure*}
    \centering
    \includegraphics[width=\textwidth]{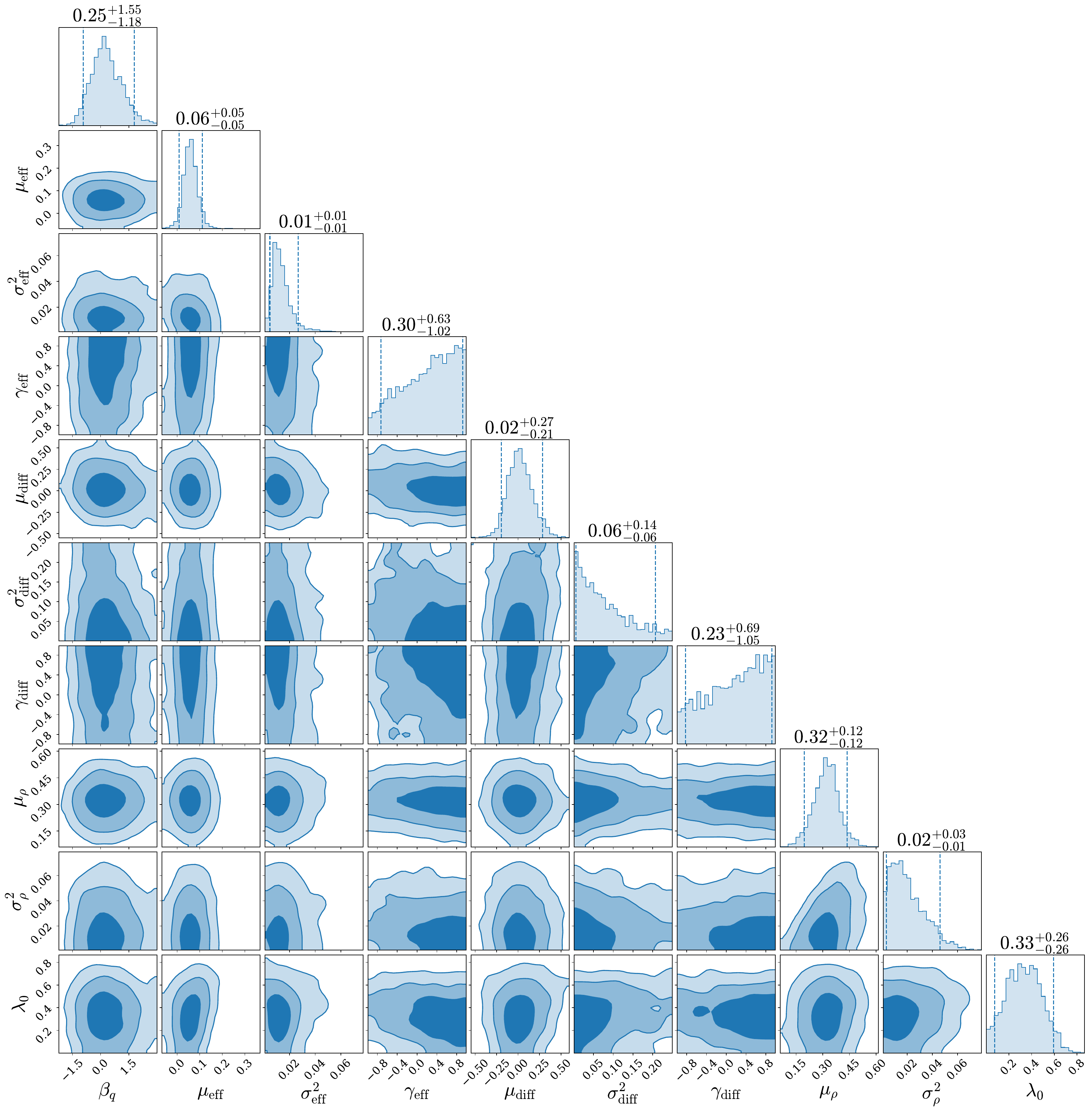}
    \caption{Posterior distributions for hyper-parameters governing the marginal distributions relevant to this study. Values listed above the posteriors are the medians, accompanied by 90\% credible intervals. These intervals are plotted as dotted lines in the one-dimensional posteriors. From darkest to lightest, the shaded regions in the two-dimensional plots represent 50, 90, and 99\% credibility regions. This plot uses results excluding GW191109.}
    \label{fig:full_corner}
\end{figure*}

In Fig.~\ref{fig:full_corner}, we plot the inferred posterior distributions of the hyper-parameters $\Lambda$, which govern the shapes of the marginal population distributions.
This includes the fraction of binaries with non-spinning black holes $\lambda_0$, which we find to be consistent with \cite{Tong_2022}.
That is, while $\lambda_0 > 0$ is preferred, the data is consistent with $\lambda_0 = 0$.

\begin{figure*}
    \centering
    \includegraphics[width=0.8\textwidth]{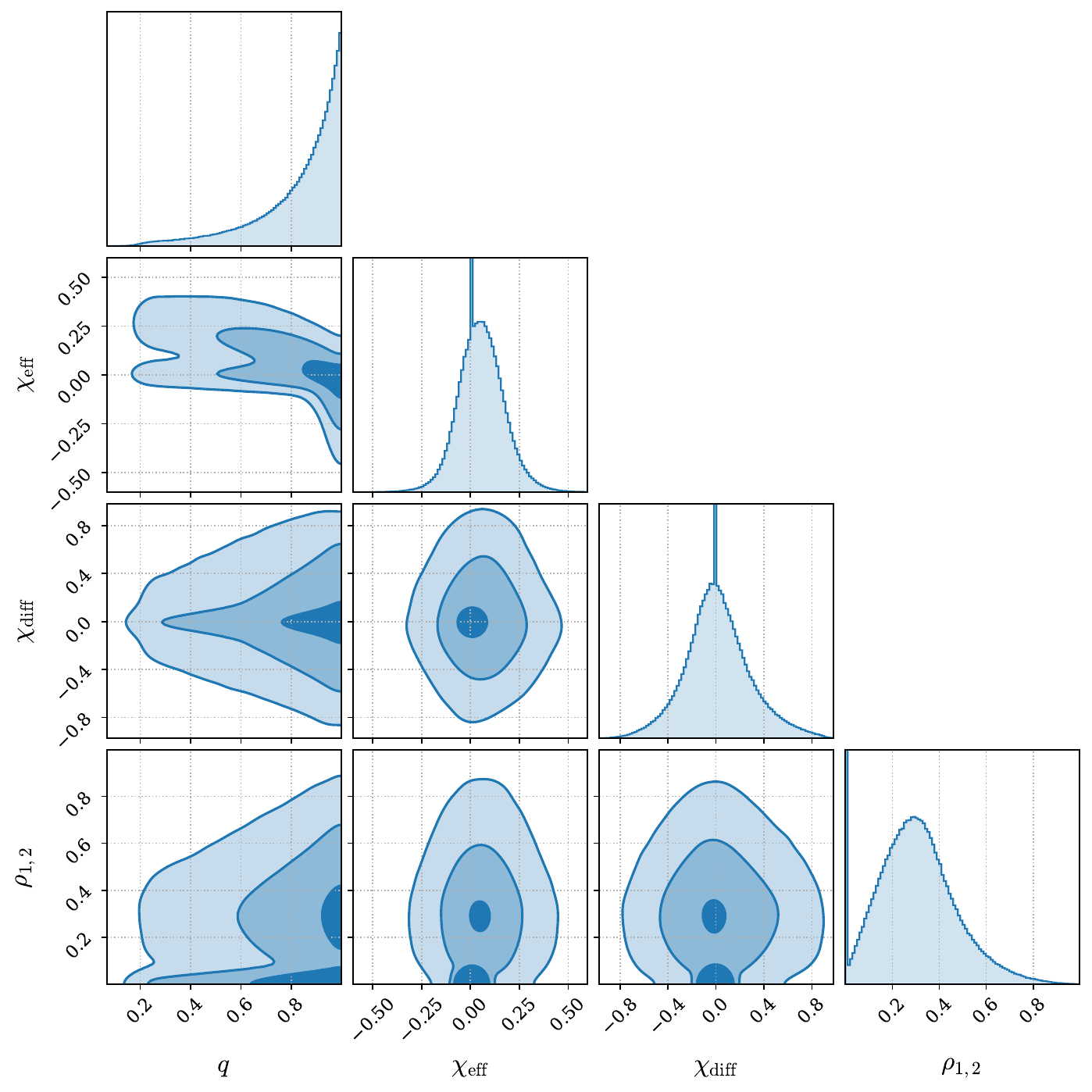}
    \caption{Population predictive distribution of mass ratio $q$, and effective spin parameters $\chieff$, $\chidif$, and $\rho_{1,2}$, reconstructed by marginalising over the hyper-parameter posterior distributions. From darkest to lightest, the shaded regions in the two-dimensional plots encapsulate 50, 90, and 99\% of the inferred BBH population. This plot uses results excluding GW191109.}
    \label{fig:full_pop}
\end{figure*}

\begin{figure*}
    \centering
    \includegraphics[width=0.95\textwidth]{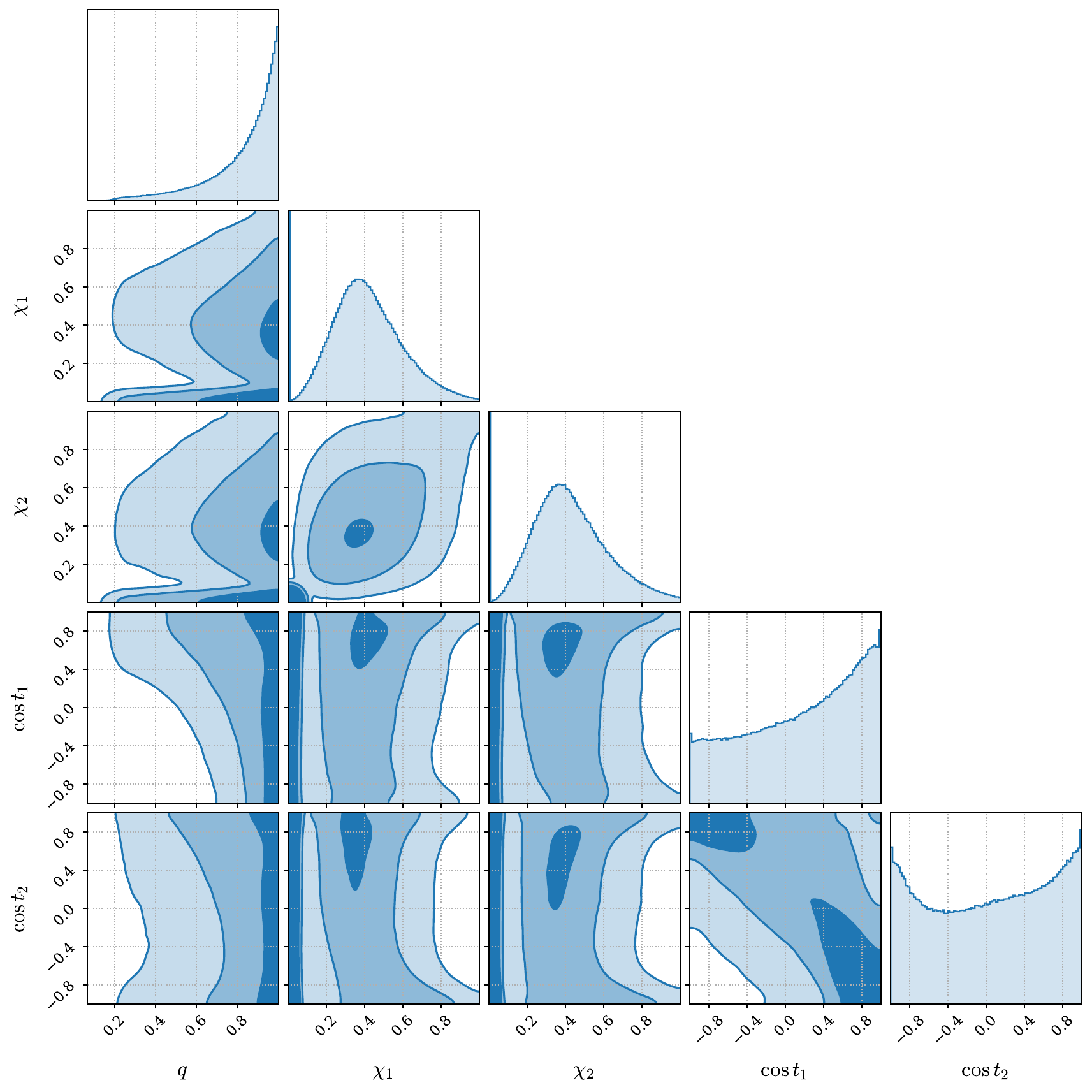}
    \caption{Implied population predictive distribution of mass ratio $q$, and physical spin parameters $\chi_{1,2}$ and $\cos t_{1,2}$, reconstructed by marginalising over the hyper-parameter posterior distributions. From darkest to lightest, the shaded regions in the two-dimensional plots encapsulate 50, 90, and 99\% of the inferred BBH population. This plot uses results excluding GW191109.
    }
    \label{fig:phy_pop}
\end{figure*}

We plot the full reconstructed population distributions of mass ratio and effective spin parameters in Fig.~\ref{fig:full_pop}, as well as the implied distributions of physical spin parameters in Fig.~\ref{fig:phy_pop}.

\bibliographystyle{aasjournal}
\bibliography{refs}

\end{document}